\documentclass[format=sigconf]{acmart}

\usepackage{hyperref}
\usepackage{natbib}
\usepackage{booktabs} 
\usepackage[utf8x]{inputenc} 
\usepackage{balance}
\usepackage{soul}

\def\BibTeX{{\rm B\kern-.05em{\sc i\kern-.025em b}\kern-.08emT\kern-.1667em\lower.7ex\hbox{E}\kern-.125emX}}
\copyrightyear{2020}
 \acmYear{2020}
 \setcopyright{acmcopyright}\acmConference[CUI '20]{2nd Conference on Conversational User Interfaces}{July 22--24, 2020}{Bilbao, Spain} \acmBooktitle{2nd Conference on Conversational User Interfaces (CUI '20), July 22--24, 2020, Bilbao, Spain}
 \acmPrice{15.00}
 \acmDOI{10.1145/3405755.3406118} \acmISBN{978-1-4503-7544-3/20/07}

\begin{document}

%
\title{Mental Workload and Language Production in Non-Native Speaker IPA Interaction}

%
\author{Yunhan Wu}
\affiliation{University College Dublin}
\email{yunhan.wu@ucdconnect.ie}

\author{Justin Edwards}
\affiliation{University College Dublin}
\email{justin.edwards@ucdconnect.ie}

\author{Orla Cooney}
\affiliation{University College Dublin}
\email{orla.cooney@ucdconnect.ie}

\author{Anna Bleakley}
\affiliation{University College Dublin}
\email{anna.bleakley@ucdconnect.ie}

\author{Philip R. Doyle}
\affiliation{University College Dublin}
\email{philip.doyle1@ucdconnect.ie}

\author{Leigh Clark}
\affiliation{Swansea University}
\email{l.m.h.clark@swansea.ac.uk}

\author{Daniel Rough}
\affiliation{University College Dublin}
\email{daniel.rough@ucd.ie}

\author{Benjamin R. Cowan}
\affiliation{University College Dublin}
\email{benjamin.cowan@ucd.ie}

%

%
\renewcommand{\shortauthors}{Wu et al.}
\begin{abstract}
Through proliferation on smartphones and smart speakers, intelligent personal assistants (IPAs) have made speech a common interaction modality. Yet, due to linguistic coverage and varying levels of functionality, many speakers engage with IPAs using a non-native language. This may impact the mental workload and pattern of language production displayed by non-native speakers. We present a mixed-design experiment, wherein native (L1) and non-native (L2) English speakers completed tasks with IPAs through smartphones and smart speakers. We found significantly higher mental workload for L2 speakers during IPA interactions. Contrary to our hypotheses, we found no significant differences between L1 and L2 speakers in terms of number of turns, lexical complexity, diversity, or lexical adaptation when encountering errors. These findings are discussed in relation to language production and processing load increases for L2 speakers in IPA interaction.     
\end{abstract}

%
%

%

%

%
\maketitle

\section{INTRODUCTION}
Intelligent personal assistants (IPAs) like Google Assistant have increased the popularity of speech as an interaction modality \cite{Clark2019}. Primarily used on smart speakers and smartphones \cite{olson_2019_2019}, these assistants can be used in a number of different languages, but coverage and functionality across these languages is not comprehensive \cite{kinsella_2019}, requiring many users to interact using a non-native language. This includes those using English as a second language, hereby referred to as L2 speakers. Interacting with IPAs in this way is likely to be significantly more challenging than the interaction experienced by those using English as their native language (L1 speakers). For instance, L2 speakers tend to experience difficulty in lexical retrieval \cite{gollan2004tot, segalowitz2005automaticity}, because of an incomplete knowledge of the language being used~\cite{Wiese1984}, with production being less automatized when compared to L1 users~\cite{dornyei1998problem}. Alongside increased demands in processing and planning utterances in a second language, this means L2 users may experience a significantly higher mental workload~\cite{Dornic1979,Wiese1984} when engaging with IPAs. These factors may also lead them to approach the interaction differently \cite{wu_see_2020, pyae_19}. Our research explores this empirically, by comparing the mental workload and language choices made by L1 and L2 speakers when interacting with IPAs across smart speakers and smartphones.

Our study identified significant differences in cognitive demand between the two speaker groups. Specifically, we found L2 speakers experience significantly higher levels of mental workload when interacting with IPAs in their non-native language compared to L1 speakers. Contrary to expectations, L1 and L2 speakers did not significantly vary in the number of commands needed to complete tasks, number of words used per command, the diversity of their lexicon, nor their levels of adaptation when they experienced errors during interaction. Our findings are the first to focus on the cognitive and linguistic aspects in L2 IPA use. 
We discuss the findings in relation to the cognitive mechanisms that may be present when interacting with IPAs as an L2 speaker.

\section{Related Work}

\subsection{Language production in speech interface interaction}
Current work on language production in speech interface interaction almost universally observes the language choices of L1 speakers. Even then, the volume of work on this topic is limited~\cite{Clark2019}, with a focus on comparing language production in interactions between human-machine and human-human interlocutors. Such work finds that users tend to vary significantly in how they interact with systems compared to how they interact with humans \cite{amalberti_user_1993}, although similar mechanisms may influence language production \cite{cowan2017they, Cowan2019}. People tend to use fewer topic shifts, use more words, as well as use fewer anaphora and fillers when interacting with computers as opposed to human partners. Similarly, people tend to use more basic lexical choices and grammatically simpler utterances \cite{bell1999repetition} when interacting with computers compared to other people \cite{kennedy_dialogue_1988}. 

This tendency to vary speech choices based on partner type is thought to be driven by the perception of a computer's competence as a dialogue partner (i.e., a user's partner model), whereby people see voice user interfaces as at-risk listeners \cite{oviatt_linguistic_1998}. This is similar to the mechanisms for adaptation proposed in psycholinguistics literature, which highlight the tendency for partners to select their language with the perceptions of their audience in mind, termed \emph{audience design} \cite{bell_language_1984}. A similar effect has recently been shown to operate on lexical choice in speech interface interaction, whereby participants interacting with a US-accented system were significantly more likely to use US lexical terms than when interacting with an Irish-accented system \cite{Cowan2019}.

\subsection{L2 speakers and speech interfaces}
Recent work comparing IPA use by both L1 and L2 speakers has focused on user experience as opposed to observing their interaction from a cognitive and linguistic perspective. L2 speakers see IPAs as more difficult to use than do L1 speakers \cite{pyae_18, pyae_19}. Recent work has also found that L2 speakers perceive difficulties in trying to use the right sentence structures or retrieving the right lexical terms \cite{wu_see_2020} when speaking to IPAs, with L2 speakers feeling they have to rephrase utterances, causing frustration \cite{pyae_19}. Research on L2 language production offers potential explanations for these perceived difficulties. It is widely acknowledged that L2 speakers tend to have an incomplete knowledge of the non-native language being used when compared to L1 speakers~\cite{Wiese1984,dornyei1998problem}. Along with a comparative lack of automatisation of the cognitive processes for language production within a second language \cite{dornyei1998problem}, this means L2 users must resort to specific production strategies to mitigate these production barriers. These include replacing lexical items, reducing message complexity or describing the meaning of words that are hard to retrieve~\cite{dornyei1998problem}. Paired with the need to process non-native speech when in dialogue, this means L2 speakers experience considerable cognitive load when having to converse in a second language~\cite{Dornic1979,Wiese1984}.

Accented speech and the need for longer planning time may also lead to L2 users experiencing difficulties in commands being understood, with the system either not recognising commands or interrupting the user before commands are complete  \cite{wu_see_2020, jain2018improved}. When they encounter communication breakdowns in IPA use, L2 speakers tend to use common strategies to repair commands such as repeating and rephrasing utterances~\cite{ipa_learning_l2}. Yet, the effective planning of error repair may depend on the type of device being used. For example, L2 speakers have emphasised the benefit of using visual feedback \cite {pyae_19}, allowing them to use further visual information (e.g., transcriptions of the conversation) to diagnose errors in their commands as well as process system prompts, making them more effective when using IPAs \cite{ipa_learning_l2, wu_see_2020}. 

\section{Research aims \& hypotheses}
Although a number of users engage with IPAs in their non-native language, research on cognitive concepts such as the mental workload and the language they produce in interaction is scant. It is therefore critical that we widen research to include the experiences of non-native speakers \cite {pyae_18, pyae_19}. Our study focuses on linguistic and cognitive aspects of L2 speaker interaction. We focus on the mental workload experienced by L2 IPA users in comparison to L1 users, while also exploring the differences in language production between the two groups when completing tasks with an IPA. 

We hypothesise that, due to planning, generating and processing speech utterances in a different language, L2 speakers are likely to experience significantly higher mental workload in IPA interaction compared to L1 speakers (H1). We also hypothesise that, due to speech recognition and planning time difficulties \cite{wu_see_2020}, L2 speakers may need significantly more turns when conducting a task than L1 speakers (H2). Due to lexical retrieval and knowledge constraints compared to L1 speakers, we also hypothesise L2 speakers will have significantly fewer words per utterance (H3), lower lexical diversity than L1 speakers in interaction (H4) and may vary in their levels of adaptation in comparison to L1 speakers when experiencing errors (H5). 

Based on work emphasising the importance of visual modalities in supporting L2 speaker IPA use \cite{wu_see_2020, pyae_19}, we also hypothesise that these effects may vary significantly by device. Specifically, the visual feedback afforded by Google Assistant on a smartphone may lead to reduced mental workload for L2 speakers due to visual output supporting error diagnosis and system query understanding (H6). As visual support helps users diagnose and correct errors, we also hypothesise that using a smartphone may  significantly affect the number of commands per turn (H7) and the number of words per command (H8), while also impacting lexical diversity (H9) and levels of adaptation (H10) for L2 speakers.

\section{Method}
To investigate these hypotheses, we designed a study that enabled us to quantitatively compare the cognitive workload and linguistic properties of L1 and L2 speakers in their interaction with IPAs. The study received ethical approval through the university's ethics procedures for low risk projects.

\subsection{Participants}
A sample of 33 participants (F=14, M=18, Prefer not to say=1) with a mean age of 28.1 years (SD=9.8 years) took part in the study. These were all recruited from students and staff at a European university via email, campus-wide posters, and snowball sampling. One participant was removed due to technical difficulties in recording their utterances, leaving 32 participants in the sample. 16 (F=8, M=7, Prefer not to say=1) were native English speakers, and 16 (F=6, M=10) were native Mandarin speakers, who used English as their non-native language. These Mandarin speakers self-reported their English proficiency as moderate (7 point Likert Scale: 1 = Not at all proficient; 7 = Extremely proficient; Mean=4.21, SD=0.7). 78.1\%  (N=25) of our sample had used IPAs before, with 9.4\% (N=3) using IPAs frequently or very frequently. For those that had used IPAs before, Siri (56\%) was most commonly used, followed by Amazon Alexa (36\%) and Google Assistant (12\%). Each participant was given a €10 voucher as an honorarium for taking part. 

\subsection{Device type}
 The study included two device conditions. Participants interacted with Google Assistant, using both a Moto G6 smartphone (\emph{Smartphone} condition) and a Google Home Mini smart speaker (\emph{Smart speaker} condition) in a within-subjects design. We selected Google Assistant because it is commonly used on both smartphones and on smart speakers \cite{olson_2019_2019}, minimising potential variation due to differences in the IPAs being used across devices. The order of device interaction was fully counterbalanced across L1 and L2 speaker groups. 

\subsection{Task}
Participants used Google Assistant to complete a total of 12 tasks (6 with each device) across the experimental session. Experimental tasks focused on 6 common IPA tasks \cite{Tawfiq2019, dubiel_survey_2018}: 1) playing music, 2) setting an alarm, 3) converting values, 4) asking for the time in a particular location, 5) controlling device volume and 6) requesting weather information. To reduce practice effects, two versions of each task were generated, creating two sets of six tasks. Each set of tasks was used in only one of the device conditions. To eliminate the influence of written tasks on user utterances, and the potential confound of written tasks increasing L2 speaker cognitive load, all tasks were delivered to participants as pictograms (see Figure \ref{fig:tasks} - all pictograms are included in supplementary material). The order of task sets were arbitrarily assigned, ensuring they were counterbalanced as much as possible across device and speaker conditions. Task order was randomised within sets for each participant.  

\begin{figure*}[t!]
    \centering
    \includegraphics[keepaspectratio, width=0.6\textwidth]{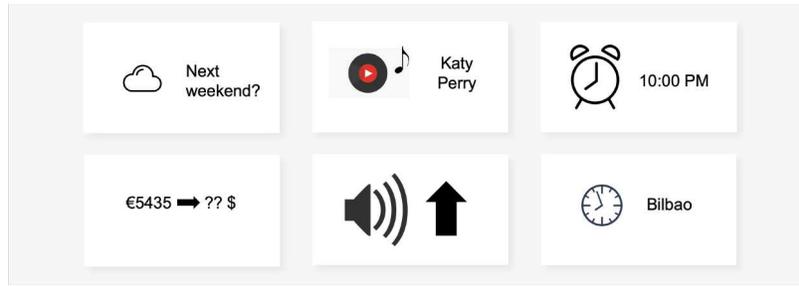}
    \caption{Example set of task pictograms}
    \label{fig:tasks}
\end{figure*}

\subsection{Measures}
\subsubsection{Mental Workload:}
To assess participants' mental workload during interaction with each of the devices, participants completed the NASA-TLX \cite{hart_development_1988}  after completing each task set. The NASA-TLX is a 6-item Likert scale (20 point scale per item) questionnaire, measuring 6 constituent factors of mental workload: \begin{itshape}Mental Demand, Physical Demand, Temporal Demand, Performance, Effort\end{itshape}, and \textit{Frustration}. Scores on the questionnaire were summed to create an overall workload (Raw TLX) score (Range: 0-120, see \cite{hart_nasa-task_2006}).

\subsubsection{Language production in interaction:}
To assess language production in interaction, user task commands were transcribed. From these transcripts, a number of measures were derived. These measures include: \textit{Number of commands per task}, \textit{Lexical complexity}, \textit{Lexical diversity per task}, \textit{Dynamic lexical adaptation}, \textit{Lexical adaption from initial command}. 

\textit{Number of commands per task} is defined as the number of utterances, starting with a wake phrase (i.e. "Hey Google" or "OK Google"), that a participant used to complete a task. 

\textit{Lexical complexity} (measured through word count per command) was derived by dividing the total word count used to complete a task by the number of turns taken. This measure represents the complexity of the utterance, and follows measures of L2 linguistic complexity used in text-based research \cite{ortega2003syntactic}. As commands to speech interfaces tend to be concise, formulaic statements \cite{gilmartin_whats_2015, kennedy_dialogue_1988}, we used word count per command rather than measuring numbers of clauses as is done in other L2 complexity research \cite{ortega2003syntactic}.  

Guiraud's index of lexical diversity \cite{guiraud_les_1954} was also calculated to identify the number of unique words used when completing a task (\textit{Lexical diversity per task}). This measure compares unique words in a command to the root of total words in a command. It is considered to be a robust alternative to diversity measures that use a direct ratio of unique words to total words, as these measures tend to inflate diversity as utterance lengths increase \cite{van2007comparing}. 

To gauge levels of lexical adaptation for tasks that required multiple utterances to complete, we measured the Guiraud index of lexical diversity for each pair of consecutive commands within a task (\textit{Dynamic lexical adaptation}). We also measured the Guiraud index of lexical diversity for each utterance paired with the first utterance of a task to determine how much participants varied their lexical choices away from their initial command (\textit{Lexical adaption from initial command}). Both measures of adaptation were used so that different styles of adaptation would be detected. For instance, participants may make a command, try a different phrasing, then return to their original phrasing. This would result in high dynamic lexical adaptation but low lexical adaptation from initial command. Participants may alternatively adapt by changing few words across many commands, resulting in low dynamic lexical adaptation but high lexical adaptation from initial command as each utterance increasingly departs from the first attempt. Using both measures allow us to detect these differences.

\subsection{Procedure}
Upon arrival, participants were welcomed by the experimenter, given an information sheet with details about the experiment and asked to give written consent to take part in the study. Participants then completed a demographic questionnaire, giving information about their age, sex, nationality, native language, experience with IPAs and speech interfaces, and their self-reported level of English proficiency. Participants were then given instructions for the study. Within these, they were asked to also look at 6 practice pictograms with the same visual structure as those in the experimental session but different in the information requested, and write what they would say to the IPA to complete the task depicted. From these responses, experimenters ensured they were interpreting the pictograms correctly before conducting the experimental tasks. They were then asked to complete a number of tasks with Google Assistant on two devices - a smartphone and a smart speaker. These tasks were displayed on a laptop, one at a time. Participants were asked to complete a task using the Assistant and once they felt they had done so, were asked to move to the next task. After completing a set of 6 tasks with one of the devices, participants then completed the NASA-TLX. This was then repeated for the next 6 tasks, wherein they interacted with Google Assistant through the other device. After finishing all tasks with both devices, participants then completed a short post-interaction interview and were then fully debriefed as to the aims of the study, and thanked for participation. To capture participant utterances, the sessions were recorded using Audacity v. 2.3.0.

\section{Results}
Out of the total of 384 tasks, 315 were successfully completed (82\%) with 14 partially completed (3.6\%) (i.e., participants completed the task but varied the information requested). 45 tasks (11.7\%) were not successfully completed, of which 24 (6.2\%) were not completed due to technical errors. Unsuccessful and technical error tasks were excluded from the dataset analysed. Before analysis, all data was screened for outliers, with these being replaced by values of the mean $\pm$ 3 SDs as suggested in \cite{field2012discovering}. Descriptive statistics for all measures included in the study are shown in Table 1.

\subsection{Mental Workload}
\begin{figure}
  \includegraphics[width=0.4\textwidth]{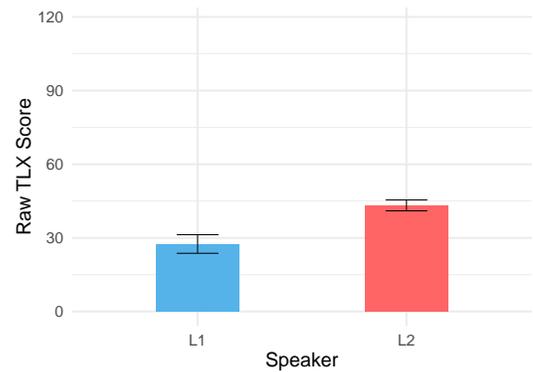}
  \caption{Mean Raw TLX scores (10\% trimmed means with trimmed standard error) for each speaker group}
  \label{fig:tlx}
\end{figure}

Due to violation of the assumption of normal distribution (p<.05), a robust mixed ANOVA with 10\% trimmed means was run using the WRS2 package (Version 1.0) \cite{mair2019} in R (Version 3.6)  \cite{rcore}. There was a statistically significant main effect of speaker on the mental workload experienced, whereby L1 speakers reported significantly lower NASA-TLX scores (Mean=27.0; SD=19.07) than L2 speakers (Mean=42.0; SD=14.37) [Q=11.74, p=.002] (see Figure \ref{fig:tlx}). This supports our first hypothesis (H1). However, there was no statistically significant main effect of device type [Q=0.28, p=.60] or interaction between speaker type and device type [Q=0.81, p=.37] on mental workload. H6 was therefore not supported. 

\subsection{Language production in interaction}
\subsubsection{Analysis Approach:}
To analyse the language production data, linear mixed-effects models (LMM) were run using the \emph{lme4} package (Version 1.1.21) \cite{bates2012package} in R (Version 3.6)  \cite{rcore}. This type of analysis allows for the modelling of fixed (i.e., device and speaker type) and random (i.e., participant and task variations) effects on specific outcomes such as lexical diversity. LMMs also allow us to model 
individual differences through random intercepts, as well as differences in how the fixed effects vary by participant and by task through modelling random slopes. We take the approach of modelling the maximal random effect structure determined by the experiment \cite{barr_random_2013}, reducing the complexity of random effects by removing higher order random slopes to facilitate convergence. We report LMM results in the text, following recent best-practice guidelines \cite{meteyard_best_2020} by also reporting all LMM analyses fully. These appear in the supplementary material. We include fixed and random effect results as well as reporting all model syntax to improve model reproducibility.

\begin{table*}[t]
\caption{Descriptive statistics by speaker and device type}  
\begin{tabular}{llll|ll|ll|ll|ll|ll}
 &  & \multicolumn{2}{|c|}{\begin{tabular}[c]{@{}c@{}}NASA-TLX\\  score\\ (10\% trimmed)\end{tabular}} & \multicolumn{2}{c|}{\begin{tabular}[c]{@{}c@{}}Number of \\ commands\\  per task\end{tabular}} & \multicolumn{2}{c|}{\begin{tabular}[c]{@{}c@{}}Lexical\\ complexity\end{tabular}} & \multicolumn{2}{c|}{\begin{tabular}[c]{@{}c@{}}Lexical \\ diversity \\ per task\end{tabular}} & \multicolumn{2}{c|}{\begin{tabular}[c]{@{}c@{}}Dynamic \\ lexical \\ adaptation\end{tabular}} & \multicolumn{2}{c}{\begin{tabular}[c]{@{}c@{}}Lexical\\ adaptation\\ from initial \\command\end{tabular}} \\ \hline 
\multicolumn{1}{l|}{Speaker} & \multicolumn{1}{l|}{Device Type} & Mean & SD & Mean & SD & Mean & SD & Mean & SD & Mean & SD & Mean & SD \\ \hline
\multicolumn{1}{l|}{L1} & \multicolumn{1}{l|}{\begin{tabular}[c]{@{}l@{}}Smart speaker\\ Smartphone\\ Total\end{tabular}} & \begin{tabular}[c]{@{}l@{}}27.36\\ 29.00\\ 27.50\end{tabular} & \begin{tabular}[c]{@{}l@{}}18.59\\ 13.38\\ 14.13\end{tabular} & \begin{tabular}[c]{@{}l@{}}2.24 \\ 2.35   \\ 2.29\end{tabular} & \begin{tabular}[c]{@{}l@{}}2.18\\ 2.14\\ 2.15\end{tabular} & \begin{tabular}[c]{@{}l@{}}8.08\\ 8.57\\ 8.32\end{tabular} & \begin{tabular}[c]{@{}l@{}}2.87\\ 3.07\\ 2.97\end{tabular} & \begin{tabular}[c]{@{}l@{}}2.61\\ 2.58\\ 2.60\end{tabular} & \begin{tabular}[c]{@{}l@{}}0.53\\ 0.58\\ 0.56\end{tabular} &  \begin{tabular}[c]{@{}l@{}}2.12\\ 2.24\\ 2.19\end{tabular} & \begin{tabular}[c]{@{}l@{}}0.67\\ 0.61\\ 0.66\end{tabular} &
\begin{tabular}[c]{@{}l@{}}0.98\\ 0.80\\ 0.88\end{tabular} & \begin{tabular}[c]{@{}l@{}}0.96\\ 0.91\\ 0.93\end{tabular} \\ \hline
\multicolumn{1}{l|}{L2} & \multicolumn{1}{l|}{\begin{tabular}[c]{@{}l@{}}Smart speaker\\ Smartphone\\ Total\end{tabular}} & \begin{tabular}[c]{@{}l@{}}47.00\\ 40.64\\ 43.23\end{tabular} & \begin{tabular}[c]{@{}l@{}}12.20\\ 8.69\\ 8.11\end{tabular} & \begin{tabular}[c]{@{}l@{}}1.84\\ 2.30\\ 2.07\end{tabular} & \begin{tabular}[c]{@{}l@{}}1.28\\ 2.02\\ 1.70\end{tabular} & \begin{tabular}[c]{@{}l@{}}7.39\\ 7.43\\ 7.42\end{tabular} & \begin{tabular}[c]{@{}l@{}}2.58\\ 2.11\\ 2.55\end{tabular} & \begin{tabular}[c]{@{}l@{}}2.45\\ 2.55\\ 2.50\end{tabular} & \begin{tabular}[c]{@{}l@{}}0.60\\ 0.53\\ 0.57\end{tabular} &  \begin{tabular}[c]{@{}l@{}}2.05\\ 1.96\\ 2.00\end{tabular} & \begin{tabular}[c]{@{}l@{}}0.66\\ 0.71\\ 0.68\end{tabular} &
\begin{tabular}[c]{@{}l@{}}0.71\\ 0.88\\ 0.80\end{tabular} & \begin{tabular}[c]{@{}l@{}}0.93\\ 0.89\\ 0.91\end{tabular} \\ \hline
\multicolumn{1}{l|}{Total} & \multicolumn{1}{l|}{\begin{tabular}[c]{@{}l@{}}Smart speaker\\ Smartphone\end{tabular}} & \begin{tabular}[c]{@{}l@{}}36.89\\ 35.27\end{tabular} & \begin{tabular}[c]{@{}l@{}}16.16\\ 10.25\end{tabular} & \begin{tabular}[c]{@{}l@{}}2.04\\ 2.33\end{tabular} & \begin{tabular}[c]{@{}l@{}}1.79\\ 2.07\end{tabular} & \begin{tabular}[c]{@{}l@{}}7.74\\ 8.01\end{tabular} & \begin{tabular}[c]{@{}l@{}}2.74\\ 2.69\end{tabular} & \begin{tabular}[c]{@{}l@{}}2.53\\ 2.57\end{tabular} & \begin{tabular}[c]{@{}l@{}}0.57\\ 0.56\end{tabular} &  \begin{tabular}[c]{@{}l@{}}2.09\\ 2.10\end{tabular} & \begin{tabular}[c]{@{}l@{}}0.66\\ 0.67\end{tabular} &
\begin{tabular}[c]{@{}l@{}}0.85\\ 0.84\end{tabular} & \begin{tabular}[c]{@{}l@{}}0.95\\ 0.90\end{tabular} 
\end{tabular}
\end{table*}

\subsubsection{Number of commands per task:}
Across the data set there was a total of 933 user commands. The LMM run showed no statistically significant effect of speaker [Unstandardized $\beta$=-0.39, SE $\beta$=0.37, 95\% CI [-1.12,0.34], t=-1.06, p=.29], device  [Unstandardized $\beta$=0.12, SE $\beta$=0.27, 95\% CI [-0.41,0.63], t=0.43, p=.67] or speaker and device interaction  [Unstandardized $\beta$=0.33, SE $\beta$=0.38, 95\% CI [-0.41,1.07], t=0.88, p=.38] on the number of user commands per task. This means that our hypotheses (H2 and H7) were not statistically supported. 

\subsubsection{Lexical complexity:}
Across the dataset there were 7112 words used to command the IPAs, with an average of 7.62 words per command. There was no statistically significant effect of speaker [Unstandardized $\beta$=-0.65, SE $\beta$=0.59, 95\% CI [-1.83,0.53], t=-1.11, p=.27], device  [Unstandardized $\beta$=0.50, SE $\beta$=0.34, 95\% CI [-0.17:1.18], t=1.45, p=.15] or speaker and device interaction  [Unstandardized $\beta$=-0.45, SE $\beta$=0.49, 95\% CI [-1.41,0.51], t=-0.92, p=.36] on the number of words used per command. Therefore our hypotheses in relation to lexical complexity (H3 and H8) were not statistically supported. 

\subsubsection{Lexical diversity per task:}
The LMM model showed no statistically significant effect of speaker type on levels of lexical diversity per task [Unstandardized $\beta$=-0.15, SE $\beta$=0.11, 95\% CI [-0.38,0.07], t=-1.38, p=.18], speaker type [Unstandardized $\beta$=-0.01, SE $\beta$=0.08, 95\% CI [-0.16,0.14] ,t=-0.19, p=.85] and speaker device interaction [Unstandardized $\beta$=0.12, SE $\beta$=0.11, 95\% CI [-0.09,0.33] ,t=1.14, p=.26]. Therefore our hypotheses in relation to lexical diversity (H4 and H9) were not statistically supported.                      

\subsubsection{Dynamic lexical adaptation:}
Over the 315 successful tasks, 116 required more than one command to complete.  Tasks that participants only used one turn to complete (N=199) were excluded from the dataset. There was no statistically significant effect of speaker [Unstandardized $\beta$=-0.04, SE $\beta$=0.16,95\% CI [-0.36,0.28], t=-0.28, p=.78], device  [Unstandardized $\beta$=0.14, SE $\beta$=0.14, 95\% CI [-0.14,0.42], t=0.98, p=.32] or speaker and device interaction  [Unstandardized $\beta$=-0.24, SE $\beta$=0.20, 95\% CI [-0.64,0.16], t=-1.20, p=.23] on the level of lexical diversity from a preceding turn. Therefore, L1 and L2 speakers did not vary in their levels of lexical adaptation from a previous utterance when having to use more than one command to complete a task. There was also no impact of device type on levels of lexical adaption from previous command, so H5 and H10 were not supported. 

\subsubsection{Lexical adaptation from initial command:}
Again, tasks where participants only used one utterance to complete the task were excluded from analysis. The LMM showed no statistically significant effect of speaker [Unstandardized $\beta$=-0.26, SE $\beta$=0.18, 95\% CI [-0.61,0.10], t=-1.43, p=.16], device [Unstandardized $\beta$=-0.17, SE $\beta$=0.17, 95\% CI [-0.51,0.17],  t=-1.01, p=.32] or speaker and device interaction [Unstandardized $\beta$=0.33, SE $\beta$=0.25, 95\% CI [-0.15,0.82], t=1.35, p=.18] on the level of lexical diversity from the first turn. It seems that both L1 and L2 speakers tend to use similar levels of lexical adaptation from their first turn, with this adaptation not being influenced by device type. This means that again H5 and H10 were not supported. 

\section{Discussion}
Our work set out to identify how using IPAs in a non-native language impacted mental workload and language production. We found L2 speakers experienced significantly higher mental workload than L1 speakers in IPA interactions across both smart speakers and smartphone devices. Although there were significant levels of workload for L2 users, there were no significant differences between L1 and L2 speakers in terms of the number of turns, words used and diversity of lexical choice. They also did not vary in the level of lexical adaptation from their initial utterances. They also did not vary in their level of lexical adaptation when comparing to a preceding turn. We discuss the interpretations of these findings below. 

\subsection{Linguistic retrieval, synthesis processing \&  workload}
Our work highlights that, even though they may show similar types of language use, L2 speakers experience significantly higher mental workload than L1 users in IPA interaction. Reasons for this are likely to involve the increased load in producing and processing utterances in a non-native language \cite{Dornic1979,dornyei1998problem}. Efforts needed for lexical retrieval in production and processing may be of particular influence. Multilingual speakers store significantly more words in their mental lexicon when compared to monolinguals, to facilitate accurate word retrieval in processing and production when using other languages. This is thought to lead to less frequent access of words across their lexicon, making activation lower and thus leading to difficulties in recall and retrieving these lexical items \cite{dornyei1998problem, segalowitz2005automaticity, gollan2004tot}. The lack of automatisation of language production processes ~\cite{dornyei1998problem}, is also likely to contribute to this load. 

In addition to production issues, many L2 speakers also find it more cognitively challenging to process and understand non-native synthetic speech  \cite{watson2013effect}. Non-native speakers find synthesis in a non-native language significantly less intelligible than do native speakers \cite{reynolds1996synthetic, mayasari_l2_synthesis, watson2013effect}. This is proposed to derive from L2 speakers' comparative unfamiliarity with their non-native language's phonological system, common syntactic structures and lexicon, which may increase cognitive load when interpreting and processing speech output \cite{watson2013effect}. In real world IPA use, this mental workload may be even higher as background noise negatively affects non-native speakers' ratings of intelligibility compared to native speakers \cite{reynolds1996synthetic}. A challenge for future HCI research is to investigate ways to mitigate this load for L2 users. 


\subsection{Lexical adaptation and limited potential for diversity}
Contrary to our hypotheses, number of commands, lexical adaptation, complexity and diversity did not vary across speaker groups or device types. There may be a number of reasons for this. Although L2 users may experience more load in lexical retrieval, IPA interaction still tends to be lexically constrained. Consequently, complex and diverse lexical choices may not be a priority, as IPAs are often seen as basic dialogue partners  \cite{clark_what_2019, branigan_role_2011, moore2017spoken}. This is contrasted by more open-ended interactions in which people have been shown to use conversational and complex linguistic structures (e.g., with automotive interfaces \cite{large_steering_2017, large2019s}). L1 and L2 speaker variance may be more stark in these types of interaction.

The opportunity for lexical variation may be further limited by the requirement to use the wake word at the start of commands, reducing the potential for variability. Additionally, although adaptation has been noted as a common strategy for error repair in human-machine dialogue \cite{kennedy_dialogue_1988, oviatt_linguistic_1998}, it may be that lexical adaptation in this instance is not the primary adaptation strategy for users. Although L2 users have suggested they may use lexical strategies in IPA use (e.g., substitution or describing the meaning of words they cannot retrieve) \cite{wu_see_2020}, adaptation of pronunciation is much more strongly emphasised by L2 speakers in previous work \cite{wu_see_2020, pyae_19}. L2 speakers tend to vary significantly from L1 speakers in other speech dimensions like tempo, rate of hesitations (e.g., filled pauses, repetitions and corrections~\cite{Wiese1984}) while also adapting syntactically or semantically~\cite{Pawley1983}. Our findings suggest that, at a lexical level, L2 speakers and L1 speakers do not vary in the limited lexical context of IPA interaction. Future work should look to explore other forms of adaptation as well as other linguistic cues in language production with IPAs across these user groups.   

\subsection{Proficiency and automaticity}
Although we found no significant difference between speakers in lexical diversity and complexity, this may be due to proficiency of the participants recruited. L2 participants rated themselves as moderately proficient and all attended an English-speaking university. These factors, together with the relative simplicity of the commands required for IPA use, may explain the lack of effect in our analysis. Increased proficiency significantly improves IPA user experience for L2 language speakers~\cite{pyae_18,pyae_19}. Increased fluency in a second language is also linked to the proceduralisation of syntactic and lexical knowledge of that language~\cite{Towell1996}. Although we found no effect in our sample, there may be differences between beginner and more advanced L2 speakers. Future work should look to identify the role that this proficiency has on language production within IPA interactions.  

\section{Limitations}
Along with L2 users being recruited from a European university where English is the primary language, all L2 users were native Mandarin speakers, which may influence the wider generalisability of results to other native and non-native language combinations. It may be that cognitive effects seen in our work vary based on similarities and differences of the languages being used, such as the phonetic or structural similarity of a non-native language to participants' native tongue. This means that L2 speakers whose native languages are more closely related to English may experience even less evident language production effects than Mandarin speakers. It is therefore important that future work explores whether similar effects are seen for L2 speakers with different native languages, as well as differing levels of language ability mentioned above. It is also important to note that future work should look to increase sample size so as to identify whether the findings are replicated across larger samples of users.  

To increase ecological validity, participants were able to control when to move on to the next task. This meant that participants could complete the tasks at their own pace and may more accurately reflect how many attempts participants are willing to give a task before abandoning it. Individual differences in this willingness are likely to influence the number of commands users made. Some were willing to try several times in order to successfully complete tasks, whereas others preferred to skip to the next task after relatively few attempts, even if they were not successful at completing the task (although we note only 5.5\% of tasks in our data were abandoned by participants).  Although the experimenters encouraged participants to try as many times as necessary, they had the freedom to move on before a successful response, which could have influenced the number of commands recorded per task. 

In relation to ecological validity, it is also important to note that our research was lab based. This allowed us to minimise potential confounds such as background noise and user distraction. Yet this context may have also made users aware that they were being recorded. Real-world IPA use is likely to vary on these dimensions in comparison to a lab based environment. Future work should therefore aim to replicate our findings in a real-world deployment.  

Rather than using text based task instructions, we used pictograms to inform participants what to complete during the study. This was to ensure that the processing of non-native language in task instructions for L2 users did not confound any mental workload effects. The use of pictograms also ensured that text-based instructions did not influence subsequent language used when making commands. Future studies with L2 speakers should investigate the mental workload and language production impact of delivering written tasks experienced by speakers in such studies.

Our findings are limited to a relatively constrained linguistic task of IPA interaction. IPAs are generally designed to perform simple tasks~\cite{cowan_what_2017,Tawfiq2019} through question-answer adjacency pair dialogues ~\cite{gilmartin_social_2017, porcheron_voice_2018}, rather than being designed for more conversational or open-ended speech tasks ~\cite{doyle_humanness_2019, clark_what_2019}. It is important that future research considers the nature of L2 speech behaviours in these more open-ended scenarios.

\section{Conclusion}
  Although IPA use has grown, fuelled by their inclusion on smart speakers and smartphones, not all languages are fully supported, leading some users to interact in a non-native language. Our study focused on these non-native (L2) speakers to understand differences in their experience of IPAs from native (L1) speakers from a cognitive and linguistic perspective. We found that L2 speakers experienced significantly higher mental workload than L1 speakers, irrespective of the device they are using. Even though they experience higher load in producing and interpreting the language from the IPA, they did not vary in the way they interacted linguistically with the IPAs, showing similar number of commands, lexical complexity, lexical diversity and lexical adaptation to L1 speakers. Our work sheds light on this under-researched set of users. CUI-based research needs to study this group in more detail to identify ways to support their IPA interactions, reducing the cognitive burden they experience.

\begin{acks}
This work was conducted with the financial support of the UCD China Scholarship Council (CSC) Scheme grant No. 201908300016, Science Foundation Ireland ADAPT Centre under Grant No. 13/RC/2106 and the Science Foundation Ireland Centre for Research Training in Digitally-Enhanced Reality (D-REAL) under Grant No. 18/CRT/6224.
\end{acks}
%
\balance
\bibliographystyle{ACM-Reference-Format}
\bibliography{cui2020.bib}

\end{document}